\documentstyle[11pt,newpasp,twoside]{article}
\def\hi{{\sc Hi}}
\def\hii{{\sc Hii}}
\def\etal{{et\thinspace al.}~}
\def\Ha{H$\alpha$}
\def\kms{{\rm\,km\ s^{-1}}}

\def\edcomment#1{\iffalse\marginpar{\raggedright\sl#1\/}\else\relax\fi}
\reversemarginpar

\begin{document}
\title{{\sc Hi} as a Probe of Structure in the Interstellar Medium 
	of External Galaxies}

\author{M. S. Oey}
\affil{Lowell Observatory, 1400 W. Mars Hill Rd., Flagstaff, AZ   86001, USA}

\begin{abstract}
This review presents a perspective on recent advances in understanding
neutral ISM structure in external galaxies.  \hi\ is a
fundamental probe of galactic baryonic material, and its structure and
distribution offer vital signatures of dynamical and evolutionary
processes that drive star formation and galaxy evolution.  New,
high-resolution \hi\ data cubes for external
galaxies now reveal the features and topology of the
entire neutral ISM, which here are considered on scales of 10 -- 1000
pc.  I focus on the two principal candidates for \hi\ structuring,
mechanical feedback from massive stars and turbulence; other
mechanisms are also considered, especially with respect to supergiant
shells.  While confirmation for both mechanical feedback and turbulent
processes exists, it remains unclear how these mechanisms yield the
global, steady-state, scale-free \hi\ properties that are observed.
Understanding the formation of filamentary structure may be key in
resolving these puzzles.  New \hi\ surveys of nearby galaxies,
combined with further theoretical studies, promise continuing
important advances.
\end{abstract}

\section{Introduction}
The distribution of neutral hydrogen in galaxies is an essential
tracer of structure in the interstellar medium (ISM), and hence, of 
dynamical and evolutionary processes that drive structure formation.
The \hi\ distribution itself, which often dominates the gas mass in
galaxies, is one of our principal probes of galactic baryonic
material.  In particular, \hi\ mapping of late-type galaxies is
well-known to reveal a gas distribution that can extend many times the
characteristic optical or stellar radius of the galaxy; vivid examples
are NGC 4449 (Hunter {\etal}1998) and NGC 6822 (de Blok \& Walter
2000), among many others.  These galaxies show massive extended disk
structure and tidal features that are unseen in other galactic
components.

I will present here an extragalactic, ``user's'' perspective on
the role of \hi\ structure as a probe of interstellar processes
relevant to galaxy evolution.
The \hi\ structure bears directly on phenomena that are fundamental
to galaxy evolution and star formation:  mechanical feedback from
massive stars and their supernovae (SNe); cloud formation and star
formation; porosity of the cool ISM, especially to ionizing radiation;
ISM phase balance and physics of phase interface regions; interstellar
mixing and chemical enrichment.  These processes are manifested in the
neutral ISM in varying ways:  superbubbles and shells presumably
result from massive star mechanical feedback; fractal structure has
been associated with turbulence; filaments result from various
processes including feedback and magnetic activity; clouds result from
gravitational effects and influence from other processes;
tidal features and spiral structure are dominated by gravitation.

Structure on scales of 10 -- 1000 pc is thought to be dominated by two
principal processes:  mechanical feedback from massive stars and
turbulence, which are discussed in turn below.  At larger spatial
scales, gravitational processes will dominate, and these will not be
considered here.  The smallest spatial scales are best studied in the
Galaxy and are discussed in these proceedings by Faison.

\section{Mechanical feedback}

At present, more is known about the effects of mechanical feedback 
on the ISM than about turbulent effects.  Supersonic winds from
massive stars, single SNe, and multiple SNe from OB associations are
well-documented to generate wind-blown bubbles, supernova remnants (SNRs),
and superbubbles in both ionized and neutral gas.  Whether the shell
is mainly ionized or neutral depends primarily on whether the parent
massive stars are still present and hot enough to ionize it, although
in some circumstances, shock ionization can also be an important
effect (e.g., Oey {\etal}2000).  The standard
model for wind-blown and SN-driven shells assumes that the
maximum available mechanical energy heats the interior of the shells
to $\sim 10^6 - 10^7$ K at an inner, reverse shock, while the outer
forward shock piles up the cool shell.  The pressure of the hot, adiabatic
interior drives the growth of the outer shell.  For continuous wind
or SN input power, this model can be described by simple
analytic relations (e.g., Ostriker \& McKee 1988).  For example, a
constant input mechanical luminosity $L$ and uniform ambient density
$n$ imply an evolution for the radius $R$, expansion velocity $v$, and
interior pressure $P$ (e.g., Weaver {\etal}1977):
\begin{equation}\label{eqR}
R\propto \bigl(L/n\bigr)^{1/5}\ t^{3/5} \quad ,
\end{equation}
\begin{equation}\label{eqV}
v\propto \bigl(L/n\bigr)^{1/5}\ t^{-2/5} \quad ,
\end{equation}
and
\begin{equation}\label{eqP}
P \propto L^{2/5}\ n^{3/5}\ t^{-4/5} \quad ,
\end{equation}
where $t$ is elapsed time.  

While clear examples of shells and superbubbles around massive stars
do exist (e.g., Cappa {\etal}1999; Oey 1996), quantitative
confirmation of this standard, adiabatic model for shell formation
is essential to understand the actual role of mechanical feedback in
interstellar processes and galaxy evolution.  Mechanical feedback is
effective over about three decades in spatial scale, and the
investigations thus use several approaches:  1.  Detailed, kinematic
tests of the adiabatic model on individual wind-blown bubbles and
superbubbles; 2. Comparison of statistical properties of superbubble
populations with model predictions; 3.  Spatial correlation of shells
with recent massive star formation; 4.  Evaluation of starburst
superwind properties with respect to the model.

\subsection{Individual shell systems}

On an individual basis, the numerous observations of ring nebulae and
superbubbles leave no doubt that these structures are generated by
mechanical feedback from massive stars.  The youngest objects are
stellar wind-dominated and ionized by the parent stars, and are easily
visible as optical shell nebulae surrounding these stars (e.g.,
Meaburn 1980; Braunsfurth \& Feitzinger 1983).  Soft X-ray
emission is found in many optical superbubbles (Chu \& Mac Low 1990; Wang \&
Helfand 1991), as is qualitatively predicted by the adiabatic
model.  Intermediate ions of {\sc C iv} and Si {\sc iv} are usually
seen in absorption through lines of sight within these young superbubbles,
presumably originating in the interface region between hot and cool
gas (Chu {\etal}1994).

\subsection{Populations of superbubbles}

As these structures age, the hot stars cool and expire, allowing
the ionized gas in the shells to recombine.  Thus, the majority of
such shell structures should be found in \hi, and most nearby
late-type galaxies that have been mapped in \hi\ do indeed show \hi\
distributions that are riddled with holes and shells.  The earliest
examples were M31 (Brinks \& Bajaja 1986), M33 (Deul \& den Hartog
1990), and Holmberg~II (Ho~II; Puche {\etal}1992).  More recently, many
late-type dwarfs have been surveyed, for example, IC~2574 and DDO~47
(Walter \& Brinks 1999, 2000).  The most revealing datasets to date
are the Australia Telescope Compact Array surveys of the Small Magellanic Cloud
(SMC; Staveley-Smith {\etal}1997) and Large Magellanic Cloud (LMC; Kim
{\etal}1998).

SNe continue to power the now-neutral shells for up to $\sim 40$ Myr,
the life expectancy of the lowest-mass core-collapse SN progenitors.
Thus, the typical stellar populations within the \hi\ shells are much
fainter and more difficult to observe than in the young, nebular
objects.  However, the statistical properties of entire \hi\ shell
populations in galaxies offer an important probe of mechanical
feedback and ISM structuring.  For example, Oey \& Clarke (1997)
derived predictions for the size distribution of superbubbles from
equations~\ref{eqR} -- \ref{eqP}.  We assumed a mechanical luminosity 
function for the OB associations,
\begin{equation}\label{eqMLF}
\phi(L)\ dL \propto L^{-\beta}\ dL \quad ,
\end{equation}
where $\beta \simeq 2$ from observed \hii\ region luminosity
functions (e.g., Kennicutt {\etal}1989).  We also assumed that the
shell growth stalls when the interior pressure (equation~\ref{eqP})
reaches that of the ambient ISM.  A few other simple assumptions yield
a steady-state differential size distribution,
\begin{equation}\label{eqsize}
N(R)\ dR \propto R^{1-2\beta}\ dR
\end{equation}
for continuous creation of the OB associations.  A value of $\beta =
2$ thus implies $N(R)\propto R^{-3}$.  This power-law distribution
extends down to the smallest shells, which correspond to individual
SNRs in this analysis.  Oey \& Clarke (1997) also derived predictions
for a constant (single-valued) mechanical luminosity function, for
a single-burst creation scenario, and for the interstellar porosities.
The porosity analysis is extended by Oey {\etal}(2001).

The SMC \hi\ shell catalog, compiled by Staveley-Smith {\etal}(1997) from
combined morphological and kinematic criteria, is perhaps the most
complete shell catalog for any galaxy to date, thanks to both the high
spatial resolution and survey sensitivity.  Number counts of \hi\
shells and \hii\ regions (Kennicutt {\etal}1989) are consistent with
the relative life expectancies of these respective objects (Oey \&
Clarke 1997), thus indicating that the \hi\ shell catalog is essentially
complete.  Significantly, the predicted and observed slopes of the
shell size distribution, respectively $-2.8\pm 0.4$ and $-2.7\pm 0.6$
are in excellent agreement, suggesting that mechanical feedback may
fully explain the bubbly \hi\ structure in the SMC.

With this intriguing result for the SMC, it is thus essential to
examine other galaxies.  Kim {\etal}(1999) found that the size 
distribution for the LMC \hi\ shells is also in good agreement with
prediction.  However, puzzlingly, the relative numbers of catalogued \hi\
shells compared to \hii\ regions in the LMC is much smaller than
expected, thus casting doubt on the completeness and evolution of the
shells.  It appears that some process is prematurely destroying the
\hi\ shells, perhaps the merging of objects due to the much higher porosity and
relative star formation in that galaxy (Oey {\etal}2001).  
Unfortunately, the significance of results
for M31, M33, and Ho~II is severely limited by the incompleteness 
of the available survey data for those galaxies, although the existing
\hi\ hole size distributions are again consistent with predictions (Oey
\& Clarke 1997).  Meanwhile,
Thilker {\etal}(1998) and Mashchenko {\etal}(1999) have developed a
shell-finding code that automatically identifies expanding shells in
\hi\ data cubes.  Their preliminary results for statistical properties
of shell populations in NGC 2403 are quite encouraging.
Application of this code to new data cubes, e.g., M33 (Thilker \& Braun,
these proceedings) should offer important insights that are
unhampered by selection effects of subjectively compiled catalogs.

Additional statistical properties of \hi\ shell populations remain to
be exploited.  Oey \& Clarke (1999) derived the differential
distribution of the expansion velocities:
\begin{equation}
N(v) \propto v^{-7/2} \ \ , \quad \beta > 1.5 \quad .
\end{equation}
Comparison with the SMC catalog again yields encouraging agreement,
with predicted and observed power-law slopes of $-3.5$ and $-2.9\pm
1.4$, respectively.  Additional datasets remain to
be examined, and statistical properties of other parameters such as
inferred $t$ and $L$ could also be similarly studied.

\subsection{Spatial correspondence with star formation}

One of the most obvious tests of the global effect of mechanical
feedback and shell formation is to identify the parent stellar
populations, or their remains, with the superbubbles.  M31 (Brinks \&
Bajaja 1986) and M33 (Deul \& den Hartog 1990) both show 
correlations of OB assocations with \hi\ holes.  However,
Ho~II shows contradictory results, based on the \hi\
hole catalog compiled by Puche {\etal}(1992).  Tongue \& Westpfahl
(1995) found that the SN rate implied by radio continuum emission is
consistent with the hole energetics in that galaxy.  However, Rhode
{\etal}(1999) carried out a direct, $BVR$ search for remnant stellar
populations within the \hi\ holes, and found little evidence for the
existence of the expected stars.  This result was then contradicted by
Stewart {\etal}(2000), who used far-UV images from the {\sl Ultraviolet
Imaging Telescope} and H$\alpha$ images to conclude that a significant
correlation between the \hi\ holes and recent star formation does
indeed support a feedback origin for the holes.

It is perhaps unsurprising that studies of Ho~II yield these confusing
results in view of that galaxy's distance of 3 Mpc.  The LMC, which
is 60 times closer, presents much better spatial resolution and
should therefore yield correspondingly less ambiguous results.  Kim
{\etal}(1999) examined the correspondence between their \hi\
shell catalog, catalogued \hii\ regions (Davies {\etal}1976), and
H$\alpha$ imaging.  Not only do they find a correspondence, but they
are also able to identify an evolutionary sequence with respect to the
relative sizes and expansion velocities.  For shells with associated 
\Ha, they find that the \hi\ radius is larger than that for the \hii,
as would be expected for objects whose parent stars are producing both
mechanical and radiative feedback.  They also find that \hi\ shells
with associated \Ha\ show higher expansion velocities than those with
only an associated OB association, and that these in turn show higher
velocities than those with neither.  This again is consistent with an
expected evolutionary sequence, as the shell expansion velocities
decrease (equation~\ref{eqV}) along with the ionizing radiation and
hot star population.  Further investigation of the Magellanic Clouds
should reveal more quantitative details of the mechanical feedback
process (Oey, Gerken, \& Walterbos, in preparation).  

\subsection{Outstanding problems}

Although the above suggest that mechanical feedback is indeed a
dominant process in creating the populations of \hi\ holes and shells,
a number of outstanding problems with this model remain.  For example,
quantitatively, it has been difficult to reconcile the adiabatic model
(e.g., equation~\ref{eqR}) with observed parameters for individual
objects.  Most shells appear to be too small for the inferred $L/n$
implied by the observed parent stars, and this problem is seen in both
Wolf-Rayet bubbles (e.g., Treffers \& Chu 1982; Garc\'\i a-Segura \&
Mac Low 1995) and young superbubbles around OB associations (e.g.,
Brown {\etal}1992; Oey 1996).

Furthermore, \hi\ imaging of the environment around three nebular LMC
superbubbles reveals highly diverse conditions, with little evidence
of any \hi\ components associated with the ionized shells (Oey
{\etal}2002).  Although the shells will presumably eventually
recombine, it is difficult to interpret the lack of obvious \hi\ holes
in the immediate environment in terms of understanding the global \hi\
structure of the ISM.  Another worry for aggregate populations of \hi\
shells are a frequently-reported positive correlation of $v$ with $R$,
contrary to that implied by equation~\ref{eqV} (e.g., Kim {\etal}1999;
Puche {\etal}1992).  These problems suggest
that, at a minimum, our evolutionary model for these objects is more
complicated than represented by the simple adiabatic model.

\section{Supergiant shells and starbursts}

The very largest documented \hi\ shells, having sizes of order 1 kpc,
emphasize some of the problems with the mechanical feedback model, and
also highlight possible alternative shell-creating mechanisms.

The existence of infalling high-velocity clouds (HVCs) suggests that
the impact of these objects could be an important contributor to 
supergiant shell populations.  This suggestion is further supported by
galactic fountain models for disk galaxies (e.g., Shapiro \& Field
1976), which are ultimately also powered by mechanical feedback in the
disk.  A number of hydrodynamical simulations of infalling HVCs
confirm that these impacts result in shell-like structures (e.g.,
Tenorio-Tagle {\etal}1986; Rand \& Stone 1996; Santill\'an
{\etal}1999). 

In addition, tidal effects, which dominate energetics and structure
formation at the largest length scales, could also create \hi\ hole
features that resemble shells.  Note that many SN-driven shells will
not exhibit expansion velocities if they have become pressure-confined
by the ambient medium, thus a lack of observed expansion velocities
cannot distinguish between the feedback model and other models.  It
has been suggested that some of the largest holes in, e.g., M33 are
simply morphologically-suggestive inter-arm regions (Deul \& den Hartog
1990).  The same may be true of the giant hole identified by
de Blok \& Walter (2000) in NGC 6822.  Simple self-gravity effects
have also produced shell- and hole-like structures in numerical
simulations (Wada {\etal}2000), although morphologically these
structures appear more filamentary than the observations.

While such alternative mechanisms for creating shell-like structures
undoubtedly contribute to the supergiant shell population, 
the conventional mechanical feedback model nevertheless also
appears to apply in many situations.  For this largest category of
shells, the required amount of star formation often may be implausible
for certain individual objects, but plausible examples do exist.
Meaburn's (1980) LMC-4 is a well-known example that is unambiguously
linked to Shapley's Constellation~III, a large, extended complex
of young stars.  Kim {\etal}(1999) are able to identify an evolutionary 
sequence for supergiant shells in the LMC, based on the location of
\Ha\ emission, which is found on the interior of the \hi\ shells in
the youngest objects, and can highlight triggered star formation at
the shell edges in older objects.  In addition, Lee \& Irwin (1997) considered
formation mechanisms for supergiant shells in the edge-on SBc galaxy
NGC 3044.  They found no evidence of HVCs, and since the galaxy is
isolated, tidal interactions are also unable to explain the supergiant
shells.  They therefore conclude that the active star formation seen
in NGC 3044 is most likely to explain its supergiant shell structures.

Starburst galaxies are well-known to exhibit clear signatures
of mechanical feedback, including soft X-ray emission (e.g., Watson
{\etal}1984; Martin \& Kennicutt 1995; 
Strickland {\etal}2000), and high-velocity outflows (Conti
{\etal}1996; Gonzalez Delgado {\etal}1998; Johnson {\etal}2000).
The \hi\ distribution of late-type dwarf galaxies also suggests
that mechanical feedback from energetic star formation displaces the
gas in these galaxies.  Simpson \& Gottesman (2000) show that, whereas
blue compact dwarf (BCD) galaxies have centrally concentrated \hi\
distributions (van Zee {\etal}1998), low surface brightness (LSB)
dwarfs show much more diffuse, ring-like distributions of \hi.  This
suggests that these dwarf galaxies may undergo burst (BCD) stages when
the gas accumulates in the center, which then disperse the gas,
leading to quiescent (LSB) phases.

It is thus apparent that mechanical feedback and also other mechanisms
form supergiant shell structures.  Presumably the respective
mechanisms will dominate under different circumstances, and these
remain to be understood.

\section{Turbulent structure}

Besides mechanical feedback, turbulence is the other major process that is
thought to structure the ISM on 10 -- 1000 pc scales (see reviews by,
e.g., Scalo 1987; Lazarian 1999).  Unambiguously
associating \hi\ structure with turbulence is difficult, however,
since the signatures of turbulence presently are not well-defined.
The current approach for confirming the widespread effect of turbulent
processes is to identify a correspondence between the observed and
predicted statistical properties of ISM structure, especially in 
the power-law, essentially scale-free, characterizations of the
neutral ISM.  This approach is somewhat limited since a wide variety
of astrophysical phenomena yield power law parameterizations, thus it
is essential to model all available parameters.  The interpretations
are also hindered by projection effects that convolve
three-dimensional distributions of spatial, density, and velocity
information into two-dimensional \hi\ intensity and velocity
distributions.  Self-absorption effects in the \hi\ line emission also
need to be considered.  Nevertheless, important recent
advances alleviate these problems, and new
extragalactic datasets that can minimize projection effects
offer vital leverage on the study of turbulence in the neutral ISM.

\subsection{Power spectra}

Empirically, the two-dimensional spatial power spectrum of \hi\
structure has been determined in the SMC (Stanimirovi\'c {\etal}1999)
and LMC (Elmegreen {\etal}2001) from the \hi\ surveys of the
Magellanic Clouds mentioned above.  These probe spatial scales of
30 -- 4000 pc, thus ranging over three orders of magnitude.  For a power
spectrum of wavenumber $k$ given by
\begin{equation}\label{eqPS}
P(k) \propto k^{-\gamma} \quad ,
\end{equation}
the SMC is well-fitted with a power-law index $\gamma = 3.04 \pm
0.02$; and for the LMC $\gamma \sim 2.7$, having a spectrum 
that steepens at the smallest spatial scales (see below).
These values are perhaps surprisingly similar, and
also compare well with $\gamma\sim 3$, measured for the Milky Way
angular power spectrum (e.g., Green 1993; Crovisier \& Dickey 1983) over 
scales of 10 -- 200 pc.  These three galaxies presumably differ
significantly in their dynamical processes:  the SMC is a small,
three-dimensional galaxy, whereas the LMC and Milky Way are spirals
that vary dramatically in size and mass.  Thus, the similarity in the
spatial power spectra is notable, and may be indicative of similar
structuring processes in these galaxies (Stanimirovi\'c {\etal}1999).

The standard reference for turbulent power spectra is the Kolmogorov
model, which predicts that for homogeneous, isotropic, incompressible,
and adiabatic turbulence, the turbulent kinetic energy cascades to
ever-smaller scales, generating a power-law in the energy spectrum:
\begin{equation}
E(k) \propto k^{-5/3} \quad .
\end{equation}
This relation derives simply from the
assumption of a constant energy transfer rate at all scales.  
The comparison of the observed 2D spatial power spectra with the
Kolmogorov theory depends on how the projection and density effects
are modeled.  For example, Goldman (2000) assumes that density
fluctuations are directly coupled to velocity fluctuations, thereby
implying an observed 1D power spectrum exponent $m = \gamma -1 \simeq 2$,
which can be directly compared to the Kolmogorov value of 5/3.  The
steeper empirical value therefore suggests progressive energy
losses that can be attributed to compressible, perhaps
shock-dominated turbulence as is plausible for the ISM.

However, more compelling confirmation of turbulence lies in isolating
the kinematic properties.  Lazarian \& Pogosyan (2000) present a powerful
new formalism for disentangling the effects of density and velocity,
by examining statistical changes induced by the binning width
of the velocity slices.  They show that density fluctuations dominate
the spectral index for thick slices, while velocity
fluctuations dominate for thin slices, and thus it is possible
to recover the relevant spectral indices in these respective regimes.
For Kolmogorov turbulence, they predict 2D
indices of 11/3 and 8/3 for thick and thin slice regimes, respectively.
This technique has been applied to the SMC (Stanimirovi\'c \& Lazarian
2001), the LMC (Elmegreen {\etal}2001), and the Milky Way (Dickey
{\etal}2001).  In the LMC and Milky Way, a transition in spectral
index between thick and thin velocity binning indeed has been
identified, thereby revealing the velocity power spectrum.  The
observed spectral indices in all three datasets are 
broadly consistent with Kolmogorov turbulence, within an additive
factor of $\sim 0.2$.  This ability to distinguish the spectrum of
velocity fluctuations is one of our most powerful probes and
confirmation of turbulence in the neutral ISM.

\subsection{Energy input and dissipation}

The conventional energy source for turbulence is mechanical feedback,
which is also thought to be a major source of structure in the ISM, as
discussed above.  The scale-free nature of the turbulent power
spectra is consistent with the dominance of energy input on the
very largest scales, that simply cascades to smaller scales.  For
example, Goldman (2000) suggests that the most recent tidal encounter
between the LMC and the Galaxy could be an important source for the
LMC, as could differential rotation.  On the other hand, Norman \&
Ferrara (1996) compute the energy source function from only mechanical
feedback, and find that feedback alone is apparently sufficient to
maintain the turbulent velocity dispersion and multi-phase pressure
support in the ISM.  For their source function, contributions at all
scales are relevant, which is consistent with the results of Oey \&
Clarke (1997) that the usual superbubble size distribution
implies equal contributions to the interstellar porosity from objects
of all sizes.  The localized nature of turbulent dissipation (see below)
also implies that multi-scale energy injection should be relevant.  In
addition, Sellwood \& Balbus (1999) suggest that MHD instabilities, in
particular Balbus-Hawley instabilities (Balbus \& Hawley 1991) may be
a dominant source for driving the lowest-velocity turbulence.  This may
set the uniform minimum \hi\ velocity dispersions, particularly
those seen beyond the star-forming disk in spiral galaxies, where
mechanical feedback cannot explain the observed turbulent velocities.

Turbulent dissipation appears to take place rapidly, the kinetic
energy decaying as $t^{-\eta}$, with roughly $\eta\sim1$.  This has
been found for both incompressible and compressible isothermal, MHD
turbulence for conditions relevant to molecular clouds (Mac Low {\etal}1998 and
references therein).  Avila-Reese \& V\'azquez-Semadeni (2001)
investigate turbulent dissipation in the large-scale, non-isothermal
ISM and find similar results, with the caveat that their simulations are 2D.
For both forced and decaying turbulence, the dissipation length scale
is therefore of order the forcing length scale or less, and thus the
dissipation time 
scale is of order the corresponding crossing time scale.  This is
consistent with the results of Kim {\etal}(1998), who find a
correspondence between high-velocity features in the LMC \hi\ data
cube and superbubble structures.  This suggests that the kinetic energy
injection indeed remains localized near the SN sources.  A final issue
raised by Avila-Reese \& V\'azquez-Semadeni (2001) and Norman \&
Ferrara (1996) is the fate of the dissipated energy:  presumably this
heats significant quantities of gas and could be important for the
dynamics of the warm ionized medium and warm neutral medium (WNM).

\section{Fractals and Filaments}

The SMC and LMC \hi\ surveys present the first characterizations of
the entire neutral ISM of individual galaxies as fractal
(Stanimirovi\'c {\etal}1999; Elmegreen {\etal}2001).  Fractal
structure has been invoked to explain scale-free properties of star
formation, and can describe properties of Galactic molecular
clouds (e.g., Elmegreen \& Falgarone 1996).  The scale-free nature of
fractal structure also suggests that it results naturally from
turbulence, which is correspondingly scale-free.  Turbulent ISM
models indeed are generally able to generate global fractal structure
(e.g., Norman \& Ferrara 1996).

If definitive physical links can be established between turbulent
processes and global observed fractal structure, the resulting simple
parameterizations would offer powerful probes and leverage for
galactic evolutionary processes and modeling.  However, while the
correspondence between turbulence and fractal structure is plausible,
the actual physical links remain to be established.  Elmegreen's
{\etal}(2001) study of the LMC is the most thorough investigation with
respect to the global \hi\ fractal structure.  As discussed above,
they measured spatial power spectra that are consistent with a
scale-free, fractal character.  However, they also emphasize that the
power spectra are insensitive to the specific morphology of the
structure, which in the LMC is highly filamentary.  Thus, the cloud
and intercloud media are not homologous in topology, although the
morphology is apparently self-similar.  Elmegreen {\etal}(2001) also
construct turbulence-based fractal simulations of the LMC \hi, but are not
able to reproduce this filamentary structure.  They also find that the
large observed variations in column density cannot be reproduced by
the model, suggesting that phase transitions and shocks significantly
modify the \hi\ structure.  Since shells and bubbles created by
mechanical feedback are known to exist in large numbers, it is also
likely that they play an important role.  Thus it appears that
turbulence is unlikely to exclusively explain the \hi\ topology in the
LMC.  If fractals alone are invoked to characterize the structure, it
will have to be in a more complex way, e.g., ``multifractal''
characterizations (Chappell \& Scalo 2001).

Understanding the properties of filamentary structure and the
formation of filaments could be key in understanding the
dominant structuring processes in the ISM.  As illustrated by the
Elmegreen {\etal}(2001) study, the existence and degree of filamentary
structure constrains the role and nature of turbulent processes.
Since filamentary structure pervades the global \hi\ topology,
understanding the various processes that generate filaments, and 
identifying those that dominate, will therefore constrain the
dominant physical processes in the multi-phase ISM.

Braun (1995, 1997) studied the \hi\ distribution at $\sim 100$ pc
resolution in 11 nearby disk galaxies spanning the entire Hubble
sequence.  This comprehensive study reveals the remarkable, ubiquitous
existence of a ``high brightness network'' of filaments (HBN), which
comprises 20 -- 85\% of the total \hi\ line flux.  Remarkably, this
HBN is strongly associated with the star-forming disks, where it
accounts for 60 -- 90\% of the total \hi\ flux, while occupying only 
15\% of the face-on covering area.  Radial gradients are seen in the
brightness temperature, as well as a correlation with later Hubble type.
The velocity dispersion of the HBN is $\sim 6\ \kms$ in the line core,
with wings extending up to $30\ \kms$.  From the narrow line core,
Braun (1997) identifies the HBN with the cool neutral medium (CNM).
These results are broadly consistent with earlier attempts by Dickey
\& Brinks (1993) to estimate the relative fractions of CNM and 
WNM in M31 and M33 by measuring the CNM absorption
through lines of sight to background continuum sources.  They obtained
40\% and 15\% CNM fractions in M31 and M33, respectively, and these
values can now be viewed in terms of the filamentary CNM morphology.
The Galactic CNM filament recently reported by Knee \& Brunt (2001)
can also be seen in this context, as can the self-absorption
studies of Galactic \hi\ structure (e.g., papers by Gibson and Dickey in
these proceedings).  The coincidence of the HBN with the
star-forming disk found by Braun (1997) highlights the relation to
star formation and the consequences of structuring processes for
galaxy evolution. 

A number of mechanisms for filament formation have been explored,
although undoubtedly many others also need to be investigated.  
As mentioned above, simulations by Wada {\etal}(2000) generate
filamentary structure from self-gravity, which is appealing in light
of the HBN observations.  Mechanical feedback, resulting in shells and
superbubbles, clearly generate filamentary structures, although it is
unclear whether these are more ordered than is observed.  The interaction
of these shells and their fragmentation is a promising source of
filamentary structure (Scalo \& Chappell 1999).  However, filaments
can also be created without direct dynamical structuring:  Lazarian \&
Pogosyan (1997) find that simple Gaussian density fields also result
in filamentary structure.  Additional work to characterize observed
filaments and model their properties will offer vital constraints on
dominant structuring processes.

\section{Conclusion}

The two primary candidate mechanisms for structuring the neutral ISM
in galaxies, mechanical feedback from massive stars and turbulence,
are both clearly significant effects.  Signatures of both processes
are confirmed in the \hi\ properties of external galaxies, and global
characterizations of the ISM in terms of these processes are
available.  However, the specifics of the physical processes by which
these mechanisms lead to the steady-state, observed global ISM properties are
lacking.  Neither mechanical feedback nor turbulence can exclusively
explain the \hi\ topology.  Given the rapid, localized turbulent
dissipation process, a global role for turbulent structuring remains
to be confirmed.  Further studies making use of \hi\ and other
datasets are therefore essential to clarify the circumstances under
which these respective mechanisms dominate, as well as to identify
additional relevant mechanisms, e.g., HVC impacts, gravitational and
magnetic effects, and gas instabilities.  This work must ultimately lead to an
integrated view of structuring and dynamics in the multi-phase ISM and
the corresponding consequences for star formation and galaxy evolution.

\acknowledgments

It is a pleasure to thank Alex Lazarian for comments on
parts of this manuscript.  I also gratefully acknowledge support from
the conference organizers and Lowell Observatory.


\end{document}